%% file: ANA-HDBS-2020-12-PAPER.tex
\pdfoutput=1
\documentclass[cernpreprint, atlasdraft=false, texlive=2023, USenglish, texmf, orcidlogo]{atlasdoc}
 
\usepackage{atlaspackage}
\usepackage{atlasbiblatex}
 
\usepackage{atlasphysics}
 
\graphicspath{{logos/}{figures/}}
 
\usepackage{ANA-HDBS-2020-12-PAPER-defs}
\usepackage{multirow}

\AtlasTitle{Search for a new pseudoscalar decaying into a pair of muons in events with a top-quark pair at $\sqrt{s} = 13$~TeV with the ATLAS detector}

\AtlasAbstract{
A search for a new pseudoscalar $a$-boson produced in events with a top-quark pair, where the $a$-boson decays into a pair of muons, is performed using $\sqrt{s} = 13$~TeV $pp$ collision data collected with the ATLAS detector at the LHC, corresponding to an integrated luminosity of $139~\mathrm{fb}^{-1}$. The search targets the final state where only one top quark decays to an electron or muon, resulting in a signature with three leptons $e\mu\mu$ and $\mu\mu\mu$. No significant excess of events above the Standard Model expectation is observed and upper limits are set on two signal models: $pp \rightarrow t\bar{t}a$ and $pp \rightarrow t\bar{t}$ with $t \rightarrow H^\pm b$, $H^\pm \rightarrow W^\pm a$, where $a\rightarrow\mu\mu$, in the mass ranges $15$~GeV~$ < m_a < 72$~GeV and $120$~GeV~$ \leq m_{H^{\pm}} \leq 160$~GeV.
}

\PreprintIdNumber{CERN-EP-2023-070}
 
\AtlasDate{\today}

\arXivId{2304.14247}
 
\HepDataRecord{ins2654723}
 
\AtlasJournal{Phys.\ Rev.\ D}
\AtlasJournalRef{Phys.\ Rev.\ D 108 (2023) 092007}
\AtlasDOI{10.1103/PhysRevD.108.092007}

\hypersetup{pdftitle={ATLAS document},pdfauthor={The ATLAS Collaboration}}
 
\begin{document}
 
\maketitle

\section{Introduction}

\begin{figure*}[!htbp]
\begin{center}
\subfloat[]{\includegraphics[width=0.26\textwidth]{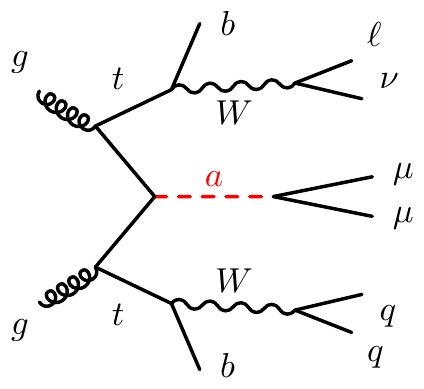}}
\hspace{2cm}
\subfloat[]{\includegraphics[width=0.26\textwidth]{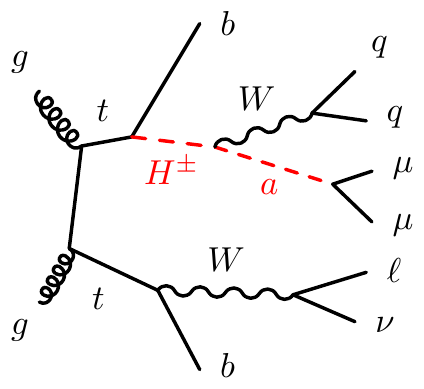}}\\
\caption{Feynman diagrams for the leading contributions to the (a) $\tta$ process and to the (b) $pp \rightarrow \ttbar$ process with $t \rightarrow H^+ b$, $H^+ \rightarrow \Wboson^+ a$.}
\label{fig:feyn_diags}
\end{center}
\end{figure*}

This paper describes a search for the production of a new light pseudoscalar particle decaying into a pair of muons in events with a top-quark pair. Such new particles are well motivated phenomenologically and have been proposed as an explanation for the excess of $\gamma$-ray emissions from the center of our galaxy~\cite{Goodenough:2009gk, Hooper:2010mq, Abazajian:2012pn, Daylan:2014rsa} in the context of Coy Dark Matter models~\cite{Boehm:2014hva,Hektor:2014kga,Arina:2014yna}. They can be colorless solutions of the naturalness problem~\cite{Chacko:2005pe,Burdman:2006tz,Burdman:2014zta,Craig:2015pha}. A light scalar can also render the electroweak phase transition strong first-order, which is one of the ingredients for electroweak baryogenesis~\cite{Profumo:2007wc,Ghorbani:2018yfr,Robens:2015gla}.
These new particles are present in several extensions of the Standard Model (SM), where new light pseudoscalars mix with fields in an extended Higgs sector, which may include additional heavy neutral and charged scalars, inheriting the Yukawa couplings to fermions. In this case, the large coupling to top quarks suggests a search for this new light pseudoscalar produced in events with a top-quark pair~\cite{Casolino:2015cza}. Two scenarios are considered: one where the new light particle $a$ is produced in association with a top-quark pair ($\ttbar a, a\rightarrow \mu\mu$) and another where a top quark decays into a new charged Higgs boson that subsequently decays into a new light particle and a $\Wboson$ boson ($t \rightarrow H^+ b$, $H^+ \rightarrow \Wboson^+ a$, $\auu$). The search focuses on the mass ranges $15~\GeV < m_a < 72$~\GeV\ and $120~\GeV \leq m_{H^{\pm}} \leq 160$~\GeV. The high mass resolution achievable for muon pairs provides a distinctive signature to search for and excellent discrimination against most of the background sources. The search targets final states with three leptons, including an electron or muon from a top-quark decay in addition to the two muons from the light pseudoscalar decay. Figure~\ref{fig:feyn_diags} shows representative Feynman diagrams for the signal processes targeted by this search.
 
This analysis uses data from proton--proton ($pp$) collisions at a center-of-mass energy of $\sqrt{s}=13~\TeV$ collected by the ATLAS experiment during \RunTwo (2015 to 2018) of the Large Hadron Collider (LHC), corresponding to an integrated luminosity of $139~\ifb$. Previous results include those from a more general search for multilepton signatures by the CMS Collaboration that sets upper limits of $1$--$10$~fb on the production cross section for $\ttbar a, a\rightarrow \mu\mu$ at 95\% confidence level in the mass ranges $m_a = 15$--$75~\GeV$ and $108$--$340~\GeV$~\cite{EXO-19-002}. The CMS Collaboration also performed a search for $H^\pm \rightarrow \Wboson^{+} a$, $\auu$ in $t \rightarrow H^\pm b$ decays targeting the mass ranges $m_{H^\pm}=120$--$160~\GeV$ and $m_a=15$--$75~\GeV$, and set upper limits of $(1.9{-}8.6)\times10^{-6}$ on the branching ratio $\text{B}(t\rightarrow bH^+,H^+\rightarrow W^+a,\auu)$ at 95\% confidence level~\cite{CMS-HIG-18-020}. Previously, the CDF Collaboration searched for $H^\pm \rightarrow \Wboson a$, $a\rightarrow\tau\tau$ in $t \rightarrow H^\pm b$ decays, targeting the mass ranges $m_{H^\pm}=90$--$160~\GeV$ and $m_a=4$--$9~\GeV$~\cite{CDF:2011pxh}.
 
\section{ATLAS detector}
The ATLAS experiment~\cite{PERF-2007-01} at the LHC is a multipurpose particle detector with a forward--backward symmetric cylindrical geometry and a near \(4\pi\) coverage in solid angle.\footnote{ATLAS uses a right-handed coordinate system with its origin at the nominal interaction point (IP) in the center of the detector and the \(z\)-axis along the beam pipe. The \(x\)-axis points from the IP to the center of the LHC ring, and the \(y\)-axis points upwards. Cylindrical coordinates \((r,\phi)\) are used in the transverse plane, \(\phi\) being the azimuthal angle around the \(z\)-axis. The pseudorapidity is defined in terms of the polar angle \(\theta\) as \(\eta = -\ln \tan(\theta/2)\). Angular distance is measured in units of \(\Delta R \equiv \sqrt{(\Delta\eta)^{2} +(\Delta\phi)^{2}}\).} It consists of an inner tracking detector (ID) surrounded by a thin superconducting solenoid providing a \qty{2}{\tesla} axial magnetic field, electromagnetic (EM) and hadron calorimeters, and a muon spectrometer. The inner tracking detector covers the pseudorapidity range \(|\eta| < 2.5\). It consists of silicon pixel, silicon microstrip, and transition radiation tracking detectors. Lead/liquid-argon (LAr) sampling calorimeters provide EM energy measurements with high granularity. A steel/scintillator-tile hadron calorimeter covers the central pseudorapidity range (\(|\eta| < 1.7\)). The endcap and forward regions are instrumented with LAr calorimeters for both the EM and hadronic energy measurements up to \(|\eta| = 4.9\). The muon spectrometer surrounds the calorimeters and is based on three large superconducting air-core toroidal magnets with eight coils each. The field integral of the toroids ranges between \num{2.0} and \qty{6.0}{\tesla\metre} across most of the detector. The muon spectrometer includes a system of precision chambers for tracking and fast detectors for triggering. A two-level trigger system is used to select events. The first-level trigger is implemented in hardware and uses a subset of the detector information to accept events at a rate below \qty{100}{\kHz}. This is followed by a software-based trigger that reduces the accepted event rate to \qty{1}{\kHz} on average depending on the data-taking conditions. An extensive software suite~\cite{ATL-SOFT-PUB-2021-001} is used in data simulation, in the reconstruction and analysis of real and simulated data, in detector operations, and in the trigger and data acquisition systems of the experiment.
 
\section{Data and simulated samples}
 
The analysis uses \pp\ collision data collected with the ATLAS detector in \RunTwo of the LHC, at a center-of-mass energy of $\sqrt{s} = 13$~\TeV, corresponding to a total integrated luminosity of $139~\mathrm{fb}^{-1}$~\cite{DAPR-2021-01}. Events were recorded using single-lepton triggers with either a low \pt\ threshold and a lepton isolation requirement, or a higher threshold but a looser identification criterion and without any isolation requirement. The lowest \pt\ threshold in the single-muon trigger was 20 (26)~\GeV~\cite{TRIG-2018-01} for data taken in 2015 (2016--2018), while in the single-electron trigger it was 24 (26)~\GeV~\cite{TRIG-2018-05}. Only events for which the LHC beams were in stable-collision mode and all relevant subsystems were operational are used~\cite{DAPR-2018-01}.
 
Simulated signal and background samples are used to describe background sources, to estimate signal efficiency and acceptance, and to design and optimize the analysis. Monte Carlo (MC) samples were produced using either the full ATLAS detector simulation~\cite{SOFT-2010-01} based on \GEANT~\cite{GEANT4:2002zbu} or a faster simulation where the full \GEANT simulation of the calorimeter response is replaced by a detailed parameterization of the shower shapes~\cite{SOFT-2010-01}. The effects of multiple inelastic interactions in the same and neighboring bunch crossings (\pileup) were modeled by overlaying each simulated hard-scattering event with inelastic $pp$ events generated with \PYTHIA[8.186]~\cite{Sjostrand:2007gs} using the \NNPDF[2.3lo] set of parton distribution functions (PDFs)~\cite{Ball:2012cx} and the A3 set of tuned parameters (tune)~\cite{ATL-PHYS-PUB-2016-017}. Simulated events are reweighted to match the \pileup conditions observed in the full \RunTwo dataset. All simulated events are processed through the same reconstruction algorithms and analysis chain as the data. Table~\ref{tab:modelling_samples} summarizes all the generated samples.
 
The precision of the matrix element (ME) generators is next-to-leading order (NLO) in quantum chromodynamics (QCD) for most samples. In all samples where the parton shower (PS), hadronization, and multiparton interactions (MPI) were generated with \PYTHIA[8], the decays of $b$- and $c$-hadrons were simulated using the \EVTGEN[1.6.0] program~\cite{EvtGen}, and the A14 tune~\cite{ATL-PHYS-PUB-2014-021} and the \NNPDF[2.3lo] PDF set~\cite{Ball:2012cx} were used.
Several samples were simulated with the \SHERPA[2.2] generator~\cite{Bothmann:2019yzt}. In this setup, NLO-accurate MEs for up to two partons, and MEs with leading-order (LO) accuracy for up to four partons, were calculated with the Comix~\cite{Gleisberg:2008fv} and \OPENLOOPS libraries. They were matched with the \SHERPA parton shower~\cite{Schumann:2007mg} by using the \MEPSatNLO prescription~\cite{Hoeche:2011fd,Hoeche:2012yf,Catani:2001cc,Hoeche:2009rj} with the tune developed by the \SHERPA authors and based on the \NNPDF[3.0nnlo] set of PDFs~\cite{Ball:2014uwa}. The top-quark mass was set to $m_t = 172.5$~\GeV. The $ttZ$ sample includes the contribution of $tt\gamma^*$ starting at $m_{\ell\ell}>5~\GeV$. The overlap between the $\Zboson$ ($\ttbar$) and the $\Zgammajets$ ($t\bar{t}\gamma$) samples is removed following the prescription in Ref.~\cite{Krause:2017nxq}.
 
\begin{table}[tp]
\renewcommand{\arraystretch}{1.2}
\setlength{\tabcolsep}{3pt}
\centering
\caption{Generator setups for samples used in this analysis.
The precision of the ME generator is NLO in QCD if no additional information is provided in parentheses. The higher-order cross section used to normalize these samples is listed in the last column if different from the one in the generator.
}
\footnotesize
\begin{tabular}{ccccc}
\toprule \midrule
Process & ME generator 		& ME PDF 					&	 PS 			& Normalization   		\\ \midrule \midrule
\multicolumn{5}{l}{Signal} \\
\midrule \midrule
$\tta$  & \MGNLO[2.7.3] 	& \NNPDF[3.0nlo] 				& \PYTHIA[8.210]		& - \\
$H^\pm \rightarrow \Wboson a$  & \MGNLO[2.3.3] (LO) 	& \NNPDF[2.3lo] 				& \PYTHIA[8.186]		& - \\
 
\midrule \midrule
\multicolumn{5}{l}{Backgrounds} \\
\midrule \midrule
 
\ttZ 	& \MGNLO[2.3.3] 	& \NNPDF[3.0nlo] 	& \PYTHIA[8.210]		& NLO QCD +NLO EW
~\cite{deFlorian:2016spz}						\\
\ttW 	& \MGNLO[2.3.3] 	& \NNPDF[3.0nlo] 				& \PYTHIA[8.210]		& NLO QCD +NLO EW
~\cite{deFlorian:2016spz}						\\
\ttH 	& \POWHEGBOX[v2] 	& \NNPDF[3.0nlo] 				& \PYTHIA[8.230]		& NLO QCD +NLO EW
~\cite{deFlorian:2016spz}					\\
$\WZjets$  	 	& \SHERPA[2.2.2] (NLO [1j], LO [3j])	& \NNPDF[3.0nnlo] 	& \SHERPA[2.2.2]		& --						\\
\tZq 	& \MGNLO[2.3.3] 	& \NNPDF[3.0nlo]     & \PYTHIA[8.230]		& --						\\
\tWZ 	& \MGNLO[2.3.3]   & \NNPDF[3.0nlo] 			& \PYTHIA[8.212]		& --						\\
$\ZZjets$  	 	& \SHERPA[2.2.2] (NLO [1j], LO [3j])  & \NNPDF[3.0nnlo] 	& \SHERPA[2.2.2]		& --						\\
$WWZ$, $WZZ$, $ZZZ$  	 	& \SHERPA[2.2.2] 				& \NNPDF[3.0nnlo] 	& \SHERPA[2.2.2]		& --						\\
$\ttbar\ttbar$  	 	& \MGNLO[2.3.3] 				& \NNPDF[3.1nlo] 	& \PYTHIA[8.230]		& NLO QCD +NLO EW~\cite{Frederix:2017wme}	\\
\ttbar	& \POWHEGBOX[v2] 	& \NNPDF[3.0nlo] 				& \PYTHIA[8.230]		& NNLO QCD + NNLL
\cite{Beneke:2011mq,Cacciari:2011hy,Baernreuther:2012ws,Czakon:2012zr,Czakon:2012pz,Czakon:2013goa,Czakon:2011xx}			\\
$\Zjets$ 	& \SHERPA[2.2.1] (NLO [2j], LO [4j])	& \NNPDF[3.0nnlo]	& \SHERPA[2.2.1]		& NNLO QCD
~\cite{Anastasiou:2003ds}						\\
$\Zgammajets$ 	& \SHERPA[2.2.8] (NLO [1j], LO [3j])	& \NNPDF[3.0nnlo]	& \SHERPA[2.2.8]		& -						\\
$\ttgamma$ 	& \MGNLO[2.3.3] (LO)	& \NNPDF[2.3lo] 				& \PYTHIA[8.212]		& -		\\
\midrule \bottomrule
\end{tabular}
 
\label{tab:modelling_samples}
\end{table}

The $t\bar{t}a$ signal samples were simulated with the \MGNLO[2.7.3] generator~\cite{Alwall:2014hca} at NLO precision. The \textsc{FeynRules}~\cite{Christensen:2008py,Alloul:2013bka} model used is based on Ref.~\cite{Bauer:2017ota} and it assumes a pseudoscalar coupling between the $a$-boson and fermions. The simulated signal samples use the \NNPDF[3.0nlo] PDF set. Tau-lepton decays of the top-quark pair are explicitly included even though their acceptance is relatively small. The production includes $gg$-, $qg$- and $\qqbar$-initiated processes. Diagrams with a single top quark, such as $tWa$ and $tqa$ are not included. The leading-order $gg$ production process is shown in Figure~\ref{fig:feyn_diags}. In the case of $\qqbar$-initiated production, three-body decays $t\to b\Wboson a$ are considered via a Breit--Wigner approximation in the whole mass spectrum of this search. Decays of the top quarks are simulated using the \MADSPIN program~\cite{Artoisenet:2012st} by setting the parameter \textsc{bwcut} to at least $(m_t-m_a)/\Gamma_t$, where $\Gamma_t$ is the top-quark width. The $a$-boson is simulated with a narrow-width approximation. The events were interfaced to \PYTHIA[8.210]~\cite{Sjostrand:2014zea}, which used the A14 tune for modeling of showering and hadronization effects. The simulated mass points correspond to $a$-boson masses of 12, 16, 20, 25, 30, 40, 50, 60, 70 and 77~\GeV.
 
The $t\to H^+b$, $H^+\to W^+a$ signal samples were simulated using the \MGNLO[2.3.3] generator interfaced with \PYTHIA[8.186], which used the A14 tune. This leading-order simulation uses the
\NNPDF[2.3lo] PDF set. Both $H^+$ and $a$ were simulated using a narrow-width assumption because the width is always smaller than the detector resolution for the investigated parameter space. The simulated signal samples include the CP-conjugate state of $H^+$. The minor contributions from single-top-quark decays are neglected.
A two-dimensional grid was chosen for the simulated $(H^+, a)$ masses from 15~\GeV\ to 75~\GeV\ in $m_a$ and from 120~\GeV\ to 160~\GeV\ in $m_{H^+}$. Mass points were generated in steps of 15~\GeV\ for the $a$-boson mass and 20~\GeV\ for the $H^+$ boson mass.
 
\section{Object and event selection}
\label{sec:selection}
Electrons are reconstructed from tracks in the ID associated with topological clusters of energy deposits in the calorimeter~\cite{EGAM-2018-01} and are required to have $\pt>27$~\GeV\ and $|\eta|<2.47$. Electrons in the calorimeter barrel--endcap transition region ($1.37 <|\eta| < 1.52$) are excluded. Electrons must satisfy the ``Medium" likelihood identification criterion and a very tight isolation requirement (``PLVTight'')~\cite{EGAM-2018-01,HIGG-2017-02}, which takes into account calorimeter energy deposits and charged-particle tracks (including the lepton track) in a cone around the lepton direction in order to reject electrons that likely originated from light- or heavy-flavor hadrons. Electron tracks must match the primary vertex of the event,\footnote{The primary vertex is the point at which a $pp$ interaction occurred. If more than one primary vertex is found, the hard-scattering primary vertex is selected as the one with the highest sum of squared transverse momenta of associated tracks~\cite{PERF-2015-01}.} i.e.\ they have to satisfy $|\zzsth|<0.5$~mm and $|\dzerosig|=|d_0/\sigma(d_0)| < 5$, where $z_0$ is the longitudinal impact parameter relative to the primary vertex and $d_0$ (with uncertainty $\sigma(d_0)$) is the transverse impact parameter relative to the beam line.
 
Muon candidates are identified by matching ID tracks to full tracks or track segments reconstructed in the muon spectrometer, using the ``Medium'' identification criterion~\cite{MUON-2018-03}. Muons are required to have $\pt>10~\GeV$ and $|\eta|<2.5$. Muon tracks must also satisfy $|\zzsth| < 0.5\,\text{mm}$ to reject \pileup tracks. A ``tight'' muon selection is defined by additionally requiring that the muon candidate satisfies $|\dzerosig|<3$ and the isolation criterion ``PFlowLoose'' with $\pT^\mu$-dependent $\Delta R$ cone radius~\cite{MUON-2018-03}. The isolation criterion is corrected for the presence of nearby muons by subtracting from the isolation sum the $\pt$ of a track associated with any other loose muon within the isolation cone of the muon. This correction is particularly important at low $m_a$, where the angular separation of the two muons from the $a$-boson decay can be small, less than the cone radius in the isolation calculation. All leptons are required to satisfy $\Delta R(\ell_1,\ell_2) > 0.1$ to reject decay chains of hadrons that produce multiple leptons.
 
Jets are reconstructed from tracks in the ID and topological energy clusters in the calorimeter~\cite{PERF-2014-07}. Tracks matched to energy clusters, as well as unmatched energy clusters, are used in a particle-flow algorithm which determines the inputs for an \antikt\ clustering algorithm~\cite{Cacciari:2008gp} with a radius parameter of 0.4. Jets are required to satisfy $\pt > 20~\GeV$ and $|\eta|<2.5$. The effect of \pileup is reduced by an algorithm which uses tracking information to require that the calorimeter-based jets are consistent with originating from the primary vertex~\cite{PERF-2014-03}. Jets containing $b$-flavored hadrons (``$b$-jets'') are identified by a multivariate discriminant ``DL1r''~\cite{FTAG-2018-01,ATLAS-CONF-2018-001,ATLAS-CONF-2018-006} combining track impact parameter values with information from secondary vertices reconstructed within the jet using a deep feed-forward neural network. A working point corresponding to 70\% efficiency for identifying $b$-jets and rejection rates of 9.4 for $c$-jets and 390 for light-flavor jets measured with simulated $\ttbar$ events is used.
 
An overlap removal procedure is applied to prevent double-counting of objects. The closest jet within $\Delta R_{y}=\sqrt{(\Delta y)^2+(\Delta \phi)^2}=0.2$ of a selected electron is removed, where $y = [(E+p_z)/(E-p_z)]/2$. If the nearest jet surviving that selection is within $\Delta R_{y} = 0.4$ of an electron, the electron is discarded. Muons are usually removed if they are separated from the nearest jet by $\Delta R_{y} < 0.4$, since this reduces the background from heavy-flavor decays inside jets.  However, if this jet has fewer than three associated tracks, the muon is kept and the jet is removed instead; this avoids an inefficiency for high-energy muons undergoing significant energy loss in the calorimeter. Electrons are removed if they share their track with a muon.
 
Events are required to have at least one primary vertex with two or more tracks with $\pt > 0.5$~\GeV~\cite{PERF-2015-01} and to have either exactly one electron and two opposite-charge muons, or three muons with total charge $\pm 1$. Events with electrons are selected with a single-electron trigger and the selected electron must match the object used in the trigger decision. Events with three muons are selected with a single-muon trigger.
 
Additional selections for muon candidates are used depending on the value of the $a$-boson mass being considered. The two opposite-charge muons with a reconstructed invariant mass closest to the mass of the hypothesized $a$-boson are referred to as ``$a$-muons''. The leading $a$-muon is required to satisfy $\pt>15~\GeV$. In events with three muons, the additional muon is interpreted as coming from a top-quark decay and is referred to as the ``top-muon''. The top-muon is required to satisfy $\pt > 27~\GeV$ and match the trigger object used in the decision. Studies with simulated signal samples show that the $a$-muons are correctly matched to muons from the $a$-boson decay in more than 98\% of the events. All muons in simulation are required to be matched to true muons originating from the prompt decays of vector bosons or $a$-bosons.
 
\section{Analysis strategy}
\label{sec:strategy}
Events are categorized into mutually exclusive regions defined by the number of muons satisfying the tight identification selection, the number of jets, the number of $b$-jets and the invariant mass of muon pairs.
 
The ``signal regions'' (SR) in both the $e\mu\mu$ and the $\mu\mu\mu$ final states are defined by $12~\GeV <m_{\mu\mu}^a<77~\GeV$, where $m_{\mu\mu}^a$ is the mass of the two $a$-muons. The range is chosen so as to minimize contamination from processes with $\Upsilon\to\mu\mu$ decays with the lower bound, and from processes with on-shell $\Zboson\to \mu\mu$ decays with the upper bound. In order to target events where only one of the top quarks decays to a lepton, the number of jets is required to be at least three and the number of $b$-jets to be at least one. In the $\mu\mu\mu$ final state, where there are two possible opposite-sign muon pairs, the mass of the ``other'' pair (not the pair chosen as $a$-muons), $m_{\mu\mu}^{\mathrm{other}}$, is required to be $m_{\mu\mu}^{\mathrm{other}}<77~\GeV$ or $m_{\mu\mu}^{\mathrm{other}}>107~\GeV$ in order to reject events with $\Zboson$ bosons. The SR regions are further binned in $m^a_{\mu\mu}$. The width of the bins is chosen to be twice the expected $m_{\mu\mu}^{a}$ resolution, defined by the Gaussian core's width in the signal model.
 
Several ``control regions'' (CR) are defined in order to target different sources of background. An ``$\onZ$'' control region is defined by requiring either $77~\GeV <m_{\mu\mu}^a<107~\GeV$ or $77~\GeV <m_{\mu\mu}^{\mathrm{other}}<107~\GeV$ but otherwise the same minimum number of jets and $b$-jets as in the SR. Two $\onZ$ CRs are defined, one in the $e\mu\mu$ final state and another one in the $\mu\mu\mu$ final state. These regions are further divided into events with 3~jets and 1~$b$-jet, events with 4~jets and 1~$b$-jet, and events with at least 4~jets and 2~$b$-jets. The regions with a small number of jets are enriched in $\Wboson\Zboson$ events, while the bins with a large number of jets are enriched in $\ttZ$ events.
 
Another CR, defined only in the $e\mu\mu$ final state, targets $\ttbar$ processes where both top quarks decay leptonically. In this case, one of the two muons is often produced in the nonprompt decay of a heavy-flavor hadron. This muon, called a ``fake muon'', is defined as the one with the same electric charge as the electron. Since both top quarks decay leptonically, events in this CR are required to have one or two jets, with exactly one of them being a $b$-jet. This region is further binned in $\pt^{\mu, \text{fake}}$.
 
In all regions described above, all muons are required to satisfy the tight selection presented in Section~\ref{sec:selection}. Events with exactly one of the selected muons failing to satisfy the tight selection criteria are assigned to one of three ``isolation sidebands'' enriched in events from background processes with nonprompt muons. The three isolation sidebands correspond to different $\pt$ ranges ($10~\GeV \leq \pt < 20~\GeV$, $20~\GeV \leq \pt < 40~\GeV$, and $\pt \geq 40~\GeV$) of the muon which does not satisfy the tight selection criteria. The binning of the isolation sidebands is the same as in the corresponding signal and control regions. The regions used in the analysis are summarized in Table~\ref{tab:regions_summary}.
 
\begin{table}[tp]
\centering \small
\renewcommand{\arraystretch}{1.4}
\setlength\tabcolsep{4pt}
\caption{Definition of the analysis regions, split according to the number of leptons, jets, and $b$-jets, and the invariant mass requirements on muon pairs.}
\begin{tabular}{lc|c|c|c|c}
\toprule
\hline
& \multicolumn{2}{c}{\textbf{Signal Regions}}  & \multicolumn{2}{c}{$\onZ$ \textbf{Control Region}} & $\ttbar$ \textbf{Control Region} \\\hline
\textbf{Channel} & $e\mu\mu$ & $\mu\mu\mu$ & $e\mu\mu$ & $\mu\mu\mu$ & $e\mu\mu$ \\
\hline
\textbf{Binning} & $m_{\mu\mu}^a$ & $m_{\mu\mu}^a$ & $n_{\text{jets}}$, $n_{b\text{-jets}}$ & $n_{\text{jets}}$, $n_{b\text{-jets}}$ & $\pt^{\mu,\text{fake}}$\\\hline
$n_{\text{electrons}}$ & 1 & 0 & 1 & 0 & 1 \\
\cline{2-6}
$n_{\text{muons}}$ & 2 & 3 & 2 & 3 & 2 \\
\cline{2-6}
\multirow{3}{*}{$m_{\mu\mu}$ [GeV]} & $12<m_{\mu\mu}^a<77$ & $12<m_{\mu\mu}^a<77$ & $77<m_{\mu\mu}^a<107$ & $77<m_{\mu\mu}^a<107$ & $12<m_{\mu\mu}^a<77$ \\
& & and & & or & \\
& - & $m_{\mu\mu}^\mathrm{other}<77$ or $>107$ & - & $77<m_{\mu\mu}^\mathrm{other}<107$ & -\\
\cline{2-6}
$n_{\text{jets}}$ &  \multicolumn{4}{c|}{$\geq3$} & $1$ or $2$ \\
\cline{2-6}
$n_{b\text{-jets}}$ &  \multicolumn{4}{c|}{$\geq1$} & $1$ \\
\hline
\bottomrule
\end{tabular}
 
\label{tab:regions_summary}
\end{table}
 
\section{Background estimation}
Several SM processes can produce final states which satisfy the object and event selections described above. The contributions from these processes are estimated with simulation or data-driven methods.
 
Background processes with all leptons originating from the prompt decay of vector bosons are described by simulation. The dominant source of background with prompt leptons is $\ttZ$, the associated production of a $\Zboson$ boson and a $\ttbar$ pair where the off-shell $\Zboson$ boson decays into a low-mass muon pair. The normalization of this process is determined by data and largely constrained by the $\onZ$ control region. The same $\ttZ$ normalization is used for all analysis channels. The categorization of the $\onZ$ CR in number of jets and $b$-jets allows a precise data-driven determination of the $\ttZ$ process normalization despite the large contamination from $\Wboson\Zboson$ events, which are subleading in the SR, as shown in Figure~\ref{fig:ttzcr}. Additional subleading sources of background in the SR include single top quarks produced in association with a $\Zboson$ boson ($t\Zboson q$ and $t\Wboson\Zboson$), as well as $\ttW$, $\ttH$, triboson $VVV$, and $\ttbar\ttbar$ processes. The normalization of each background process with prompt leptons other than $\ttZ$ is determined from simulation.
 
\begin{figure*}[!htbp]
\begin{center}
\subfloat[]{\includegraphics[width=0.5\textwidth]{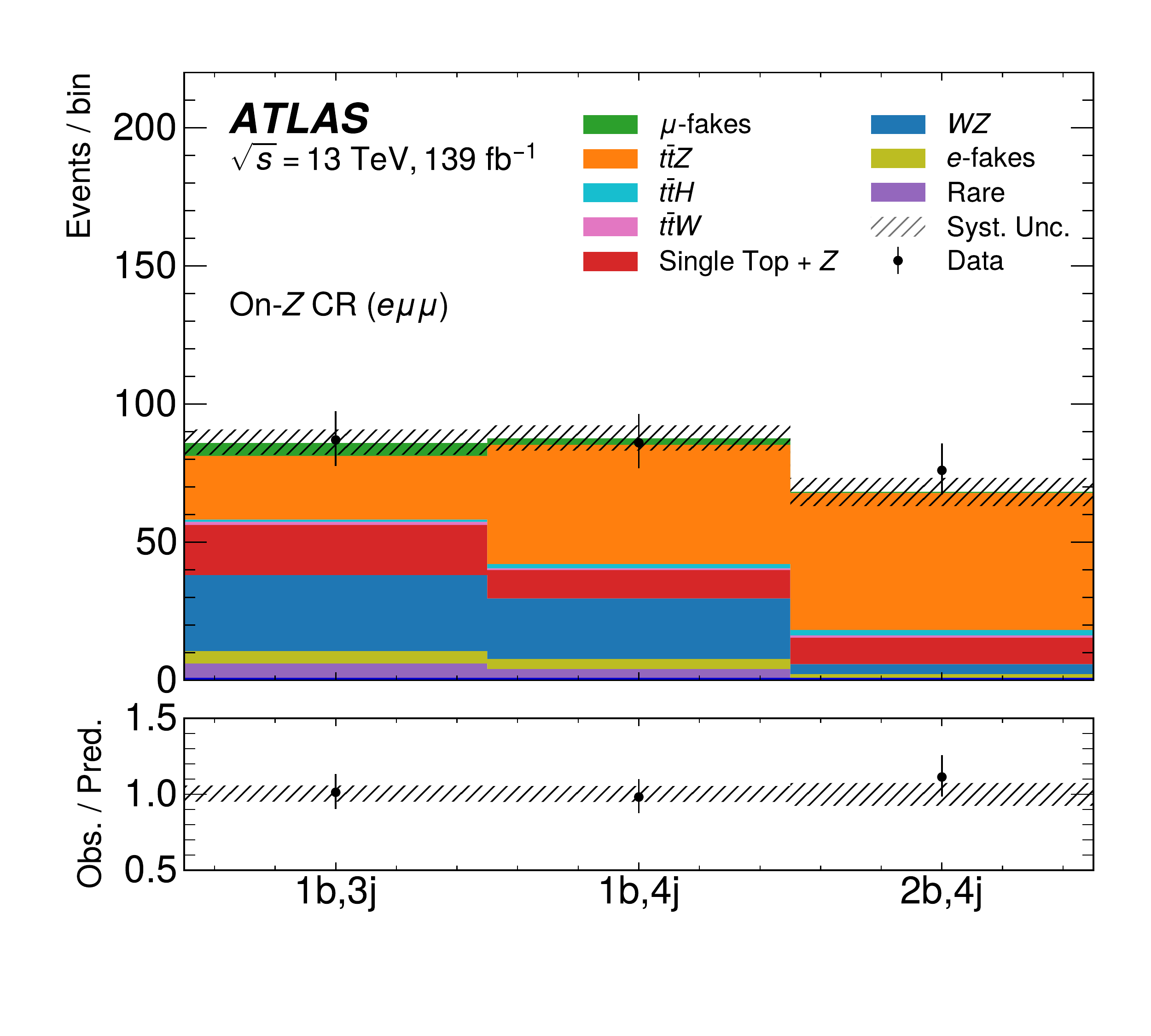}}
\subfloat[]{\includegraphics[width=0.5\textwidth]{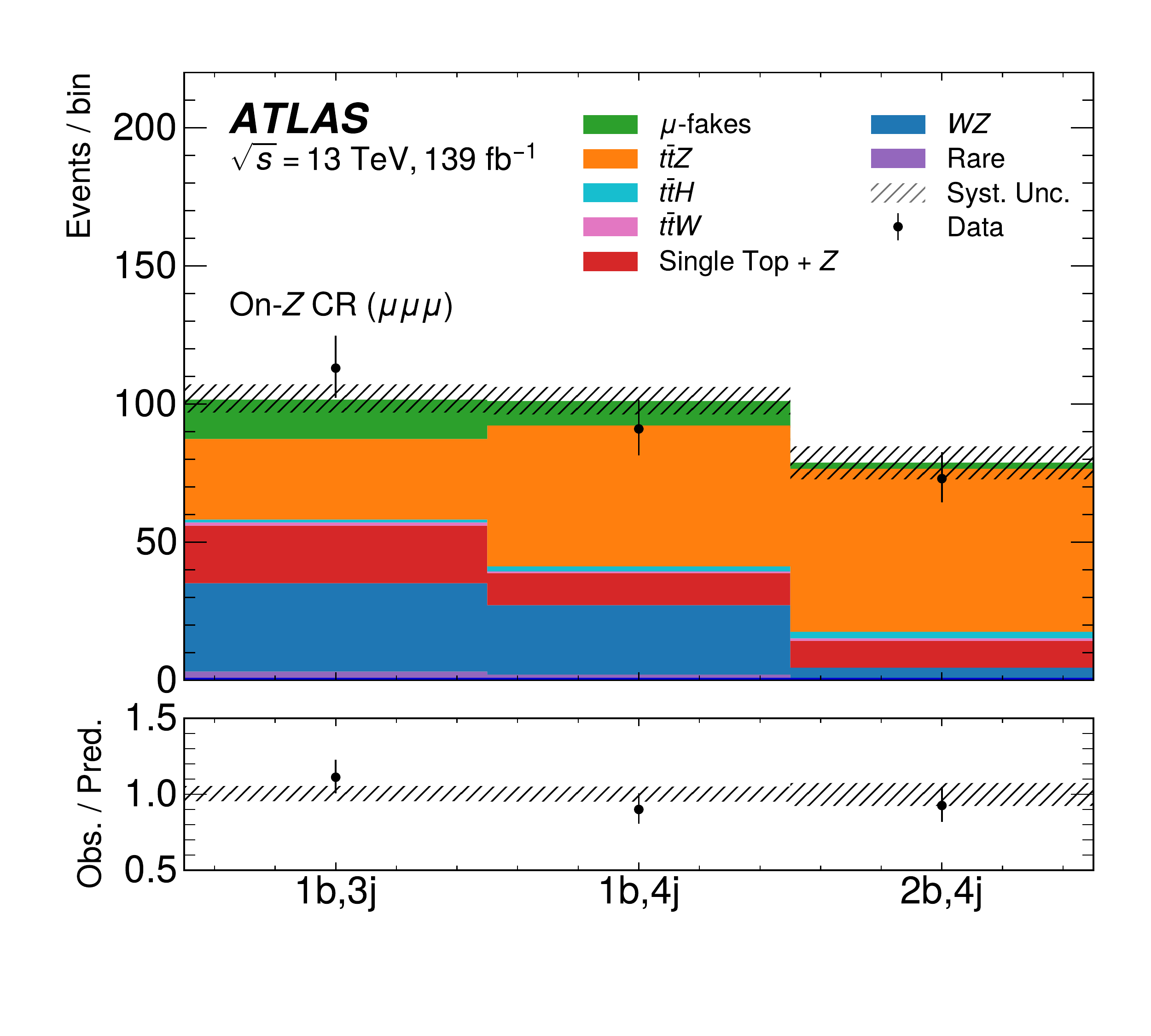}}\\
\caption{Comparison between data and expected background for the on-$\Zboson$ control region in the (a) $\euu$ and (b) $\uuu$ final states. The bins correspond to different jet and $b$-jet multiplicities. Rare background processes include $\ZZjets$, $WWZ$, $WZZ$, $ZZZ$, and $\ttbar\ttbar$. The yields correspond to the values obtained under the background-only hypothesis with the profile likelihood method described in Section~\ref{sec:result}.}
\label{fig:ttzcr}
\end{center}
\end{figure*}
 
Background processes with at least one lepton originating from the nonprompt decay of a hadron, from photon conversion, or from the misidentification of other particles, are described by simulation if the nonprompt lepton is an electron or by a data-driven method if the nonprompt lepton is a muon. Background processes with nonprompt electrons are greatly suppressed by the PLVTight isolation requirement. Nonprompt electrons are further classified depending on whether they originate from photon conversions or from hadrons. Background processes with nonprompt muons provide the largest number of events in the signal regions and the data-driven estimate ensures an accurate description. Simulation studies show that the contribution from processes with more than one nonprompt lepton is negligible.
 
As described in Section~\ref{sec:strategy}, each channel of the analysis is accompanied by three isolation sidebands enriched in events with nonprompt muons. The three sidebands are distinguished by the $\pt$ of the muon failing the tight selection criteria ($10~\GeV \leq \pt < 20~\GeV$, $20~\GeV \leq \pt < 40~\GeV$, and $\pt > 40~\GeV$). The number of events from processes with nonprompt muons in the isolation sidebands is determined by data. The number of events from processes with nonprompt muons in signal and control regions is assumed to be proportional to the number of events in the three sidebands. The proportionality constant, called the fake factor, is different for each of the three sidebands, but otherwise common to all signal and control regions. The fake factors are determined in data and are largely constrained by the $\ttbar$ control region, which is divided into the same three bins of $\pt^{\mu,\text{fake}}$. Simulation studies show that background events with nonprompt muons arise mostly ($\approx 90\%$) from $\ttbar$ processes where both top quarks decay leptonically. Simulation studies also show that nearly all other background events with nonprompt muons come from processes with semileptonic decays of heavy-flavor hadrons, akin to $\ttbar$ events.
 
Figure~\ref{fig:Unblinded_SR} shows the background composition of the two signal regions. Background processes with one nonprompt muon are called $\mu$-fakes and processes with one nonprompt electrons are called $e$-fakes. Rare background processes with prompt leptons are shown together.
 
\begin{figure*}[!htbp]
\begin{center}
\subfloat[]{\includegraphics[width=0.5\textwidth]{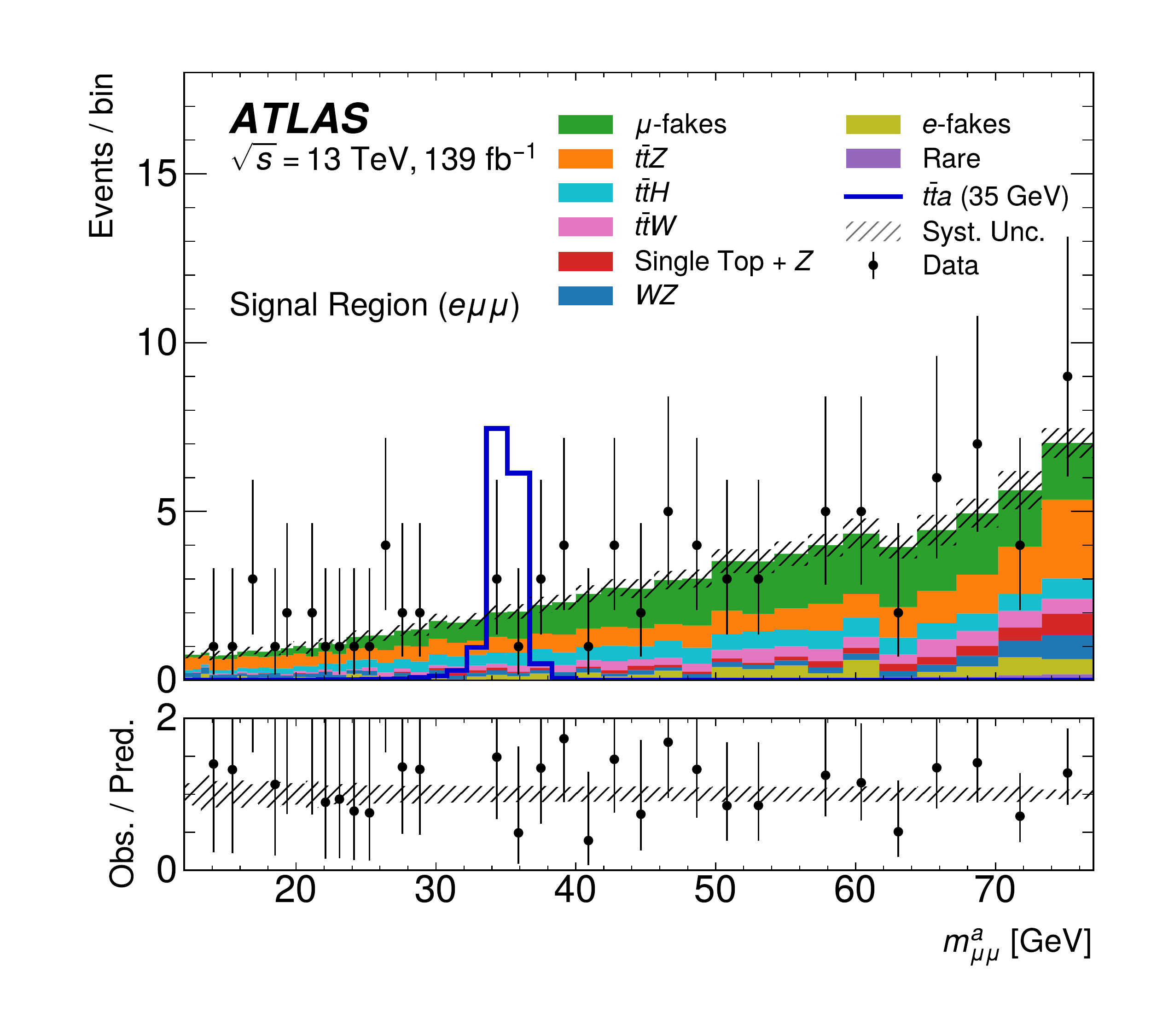}}
\subfloat[]{\includegraphics[width=0.5\textwidth]{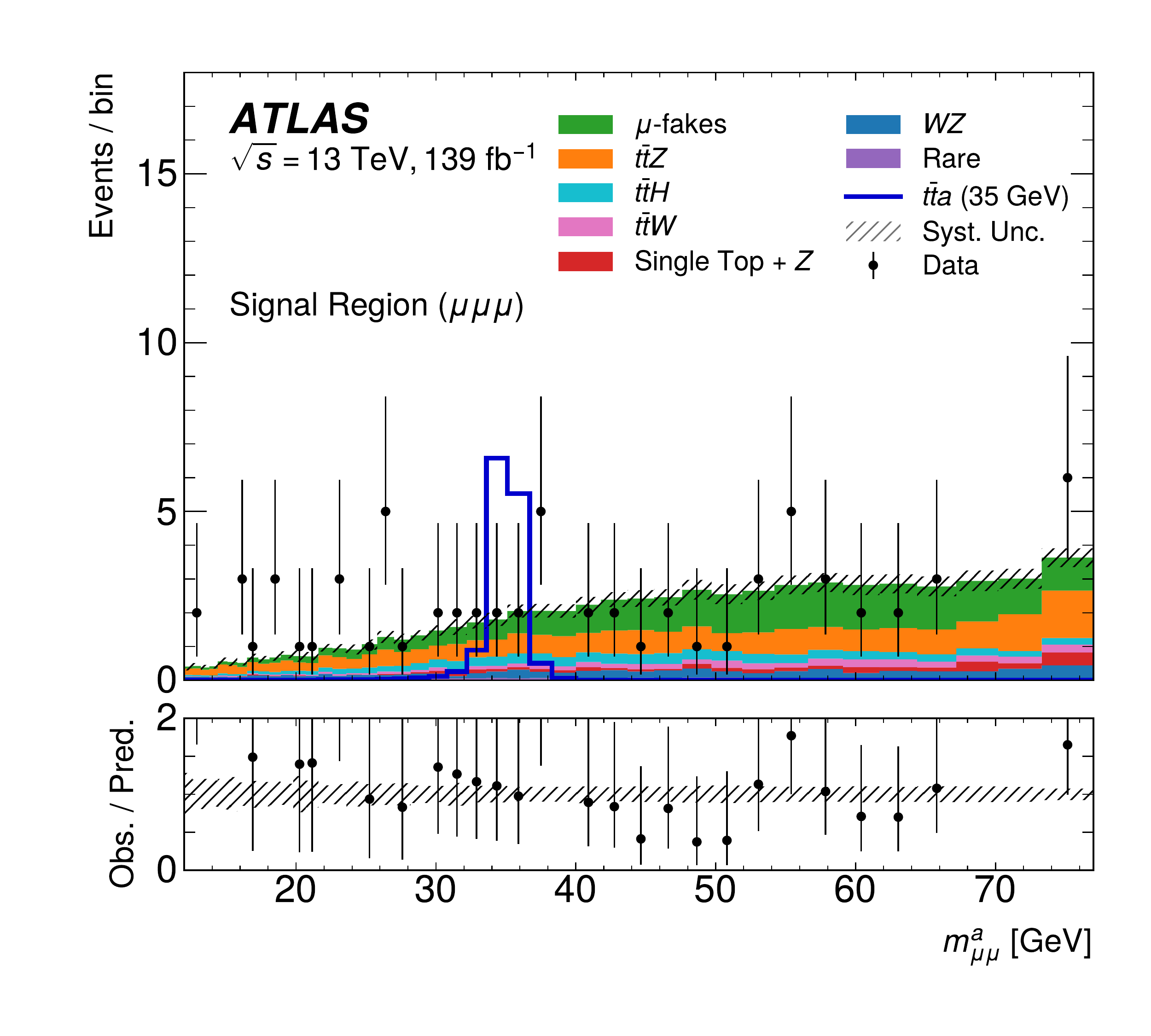}}\\
\caption{Dimuon mass distributions for data and expected background in the (a) \euu and (b) \uuu signal regions for the $m_a=35~\GeV$ hypothesis. The expected signal distribution is shown assuming $\sigma(\tta)\times \text{B}(\auu) = 4~\fb$. Rare background processes include $\ZZjets$, $WWZ$, $WZZ$, $ZZZ$, and $\ttbar\ttbar$. The yields correspond to the values obtained under the background-only hypothesis with the profile likelihood method described in Section~\ref{sec:result}.}
\label{fig:Unblinded_SR}
\end{center}
\end{figure*}
 
\section{Signal modeling}
Simulated samples of the $t\bar{t}a$, $a\to \mu\mu$ process were generated for 10 different values of the $a$-boson mass in the range between 12~\GeV\ and 77~\GeV. Similarly, in the case of the $t\to H^+b$, $H^+\to W^+a$ signal, samples for up to five different values of the $a$-boson mass were generated for three different values of $m_{H^+}$. The number of different mass hypotheses simulated is limited by the computational resources available. The gap between simulated mass hypotheses is, however, much wider than the dimuon mass resolution of the ATLAS detector.
 
A parameterized model for the $m_{\mu\mu}^a$ spectrum, where $m_{\mu\mu}^a$ is the mass of the two $a$-muons in each event, is used to probe $a$-boson mass hypotheses between the values chosen for the simulated samples. The model uses the double Crystal Ball (dCB) probability density function~\cite{Oreglia:1980cs}. The parameters of the dCB distribution are evaluated independently for each simulated $a$-boson mass using a maximum-likelihood fit. Only the Gaussian core's mean and width are observed to vary significantly, while the parameters describing the power-law tails are consistent for all $a$-boson masses in the range considered in this search. The mean and the width of the Gaussian core vary linearly with the value of the $a$-boson mass. The signal acceptance times efficiency is also observed to vary linearly with the value of the $a$-boson mass. The dCB distribution parameters, their linear dependency on the $a$-boson mass, and the model of the signal acceptance times efficiency are determined separately in the $e\mu\mu$ and $\mu\mu\mu$ channels and for the two signal models $\tta$ and $t\rightarrow H^+b$, $H^+\rightarrow \Wboson^+ a$ in order to account for the different muon kinematics in the two models. Figure~\ref{fig:signal_model} shows the dependency of the dCB width on the value of the $a$-boson mass in the $e\mu\mu$ channel. It also shows a comparison of the model obtained with the prescription described above and the distribution of events obtained from an independent simulated sample. Similar tests show excellent modeling throughout the whole mass spectrum considered in this search. Once validated, a model built with all the simulated signal mass points was used in the analysis.
 
\begin{figure*}[!htbp]
\begin{center}
\subfloat[]{\includegraphics[width=0.38\textwidth]{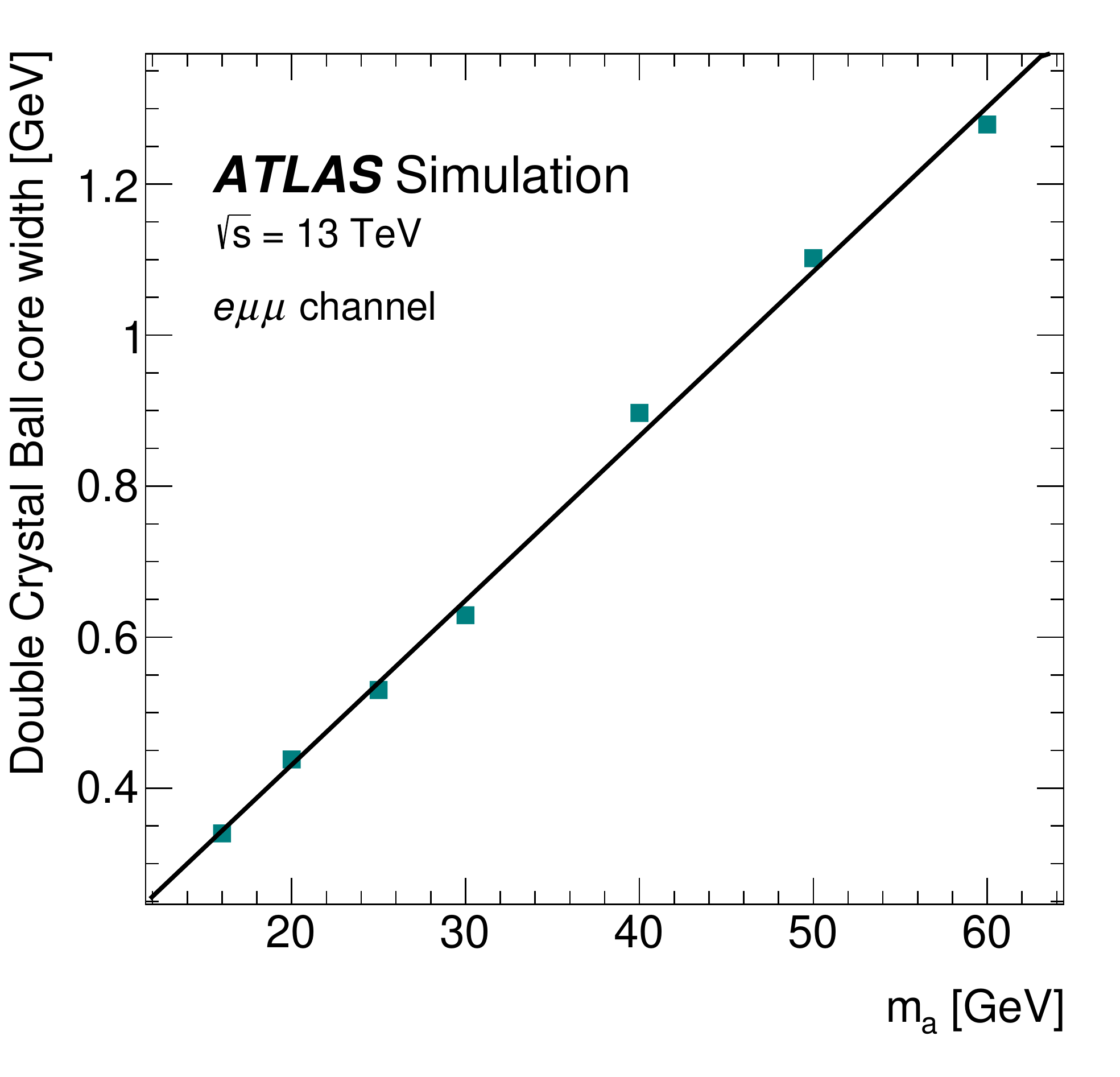}}
\hspace{1cm}
\subfloat[]{\includegraphics[width=0.38\textwidth]{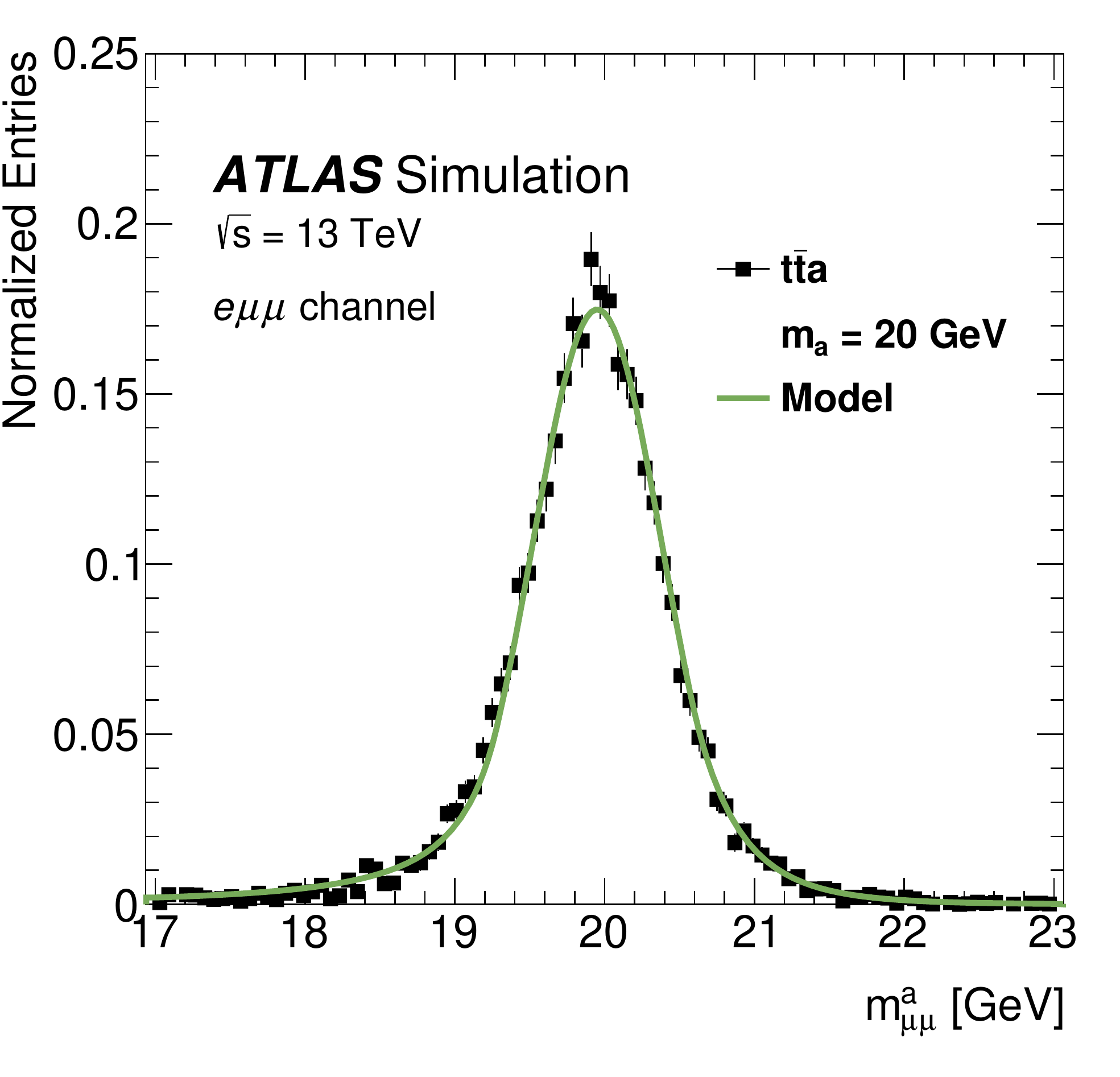}}\\
\caption{(a) Fit of the dCB width as a function of the $a$-boson mass in the $e\mu\mu$ channel. (b) Example fit of the signal using the dCB fit model compared with an independent simulated sample for the $\tta$ signal with $m_a=20~\GeV$ in the $e\mu\mu$ channel.}
\label{fig:signal_model}
\end{center}
\end{figure*}
 
\section{Systematic uncertainties}
Several sources of experimental and modeling systematic uncertainty affecting the signal acceptance and efficiency and the background normalization are considered. The uncertainty in the combined 2015--2018 integrated luminosity is 1.7\%~\cite{ATL-DAPR-PUB-2021-001}, obtained using the LUCID-2 detector~\cite{LUCID2} for the primary luminosity measurements, complemented by measurements using the inner detector and calorimeters. The uncertainty in the number of \pileup events is determined by comparing estimates based on the total inelastic cross section measured using either the luminosity detectors or the ID.
 
Uncertainties in the muon and electron efficiency, momentum scale and resolution are determined from alternative simulations, and from tag-and-probe measurements using $\Zboson\to\ell\ell$ and $J/\psi\to \ell\ell$ processes~\cite{MUON-2018-03,EGAM-2018-01}. The most important lepton uncertainties in this search are the ones related to the muon identification efficiency. These uncertainties are below 1\% in the momentum range considered.
 
Uncertainties associated with the jet energy scale are evaluated by combining information from test-beam data, LHC collision data and simulation, and the jet energy resolution uncertainty is obtained from measurements of dijet \pt-balance in data and simulation~\cite{JETM-2018-05}. Additional considerations related to jet flavor, \pileup corrections, $\eta$ dependence and high-\pt jets are included. The efficiency to identify and remove jets from \pileup is measured with $Z (\to \mu^+\mu^-)\text{\,+\,jets}$ events in data using techniques similar to those used in Ref.~\cite{PERF-2014-03}. The uncertainty in tagging $b$-jets is 2\%--10\% depending on the jet \pt. The uncertainty in mistagging $c$-jets (light jets) is 10\%--25\% (15\%--50\%) depending on the jet \pt~\cite{FTAG-2018-01,ATLAS-CONF-2018-001,ATLAS-CONF-2018-006}.
 
Uncertainties in modeling the background are assessed primarily through variations of the renormalization and factorization scales. The effects of independent variations of the two scales between twice and half their nominal values are considered and their envelope is taken as representative of uncertainties associated with missing higher-order terms in the calculation of the hard-scatter ME. For the leading $\ttZ$ background process, the scale uncertainties are approximately 5\%. Uncertainties associated with the showering and hadronization modeling are assessed through alternative \PYTHIA[8] A14 tunes. Uncertainties in the electroweak production of a single top quark in association with a $\Zboson$ boson are estimated by comparing different prescriptions to handle the interference with the $\ttZ$ process, and are found to have a negligible impact in this search.
 
Uncertainties associated with the signal processes are found to be largely independent of the $a$-boson mass hypothesis, and a single uncertainty model is used independently of the signal hypothesis being tested. Uncertainties in modeling the signal acceptance are assessed by varying the renormalization and factorization scales, as described above, as well as through comparison of simulations using either \PYTHIA[8] or \HERWIG[7.1]~\cite{Bahr:2008pv,Bellm:2015jjp} for showering and hadronization. These uncertainties are approximately 7\% for the cross section and 1\% for the acceptance. Uncertainties stemming from variations of the proton PDFs are considered for both the signal and background simulations following the NNPDF prescription~\cite{Ball:2012cx}, and are found to have negligible impact on the result.
 
The systematic uncertainties are much smaller than the data statistical uncertainty. The two largest systematic uncertainties are those related to the muon identification efficiency and the modeling of the $\ttZ$ background. These uncertainties would have an impact of 2\% and 1\%, respectively, in the measurement of a hypothetical signal cross section, compared to a statistical uncertainty of approximately 40\%--70\%, depending on the $a$-boson mass.
 
\section{Results}
\label{sec:result}
 
The presence of a signal consistent with the production of $t\bar{t}a$ or $t\to H^+b, H^+\to W^+a$, with $a\to\mu\mu$ is tested by comparing the expected background with data in narrow bins of the reconstructed $m_{\mu\mu}^{a}$. A total of 43 bins between 12~\GeV\ and 77~\GeV\ are used. For simplicity, the same binning is used in both the $e\mu\mu$ and $\mu\mu\mu$ regions. A bin width proportional to the expected width of a detected signal, illustrated in Figure~\ref{fig:signal_model}(b), ensures that any signal contribution is highly concentrated in a few bins, while keeping the analysis largely independent of the specific signal mass distribution.
 
A test statistic is built from a profile likelihood ratio $\tilde{q}_\mu=-2\ln({\cal L}(\mu,\hat{\hat{\theta}}_\mu)/{\cal L}(\hat{\mu},\hat{\theta}))$, where the likelihood ${\cal L}$ is built starting from the product of Poisson distributions in each analysis channel. The values of the parameters that maximize the likelihood function are $\hat{\mu}$ and $\hat{\theta}$, and the values of the nuisance parameters that maximize the likelihood function for a given value of $\mu$ are $\hat{\hat{\theta}}_\mu$~\cite{Cowan:2010js}. The test statistic value is set to zero if the hypothesis $\mu$ is lower than the maximum-likelihood estimator $\hat{\mu}$. The expected number of events is parameterized as a function of the signal strength $\mu$, which multiplies a reference signal cross section of $1\,\fb$, and a large number of nuisance parameters $\theta$ describing the background model and systematic uncertainties. The parameters describing the background model include the ones describing the overall $\ttZ$ normalization, the background yield from processes with nonprompt leptons in three isolation sidebands independently for each analysis channel, and the three global fake factors. The analysis channels include the 43 bins in each signal region, but also the regions with various numbers of jets and $b$-jets in the $\onZ$ control region, and the three $\pt^{\mu,\text{fake}}$ regions of the $\ttbar$ control region. Hypothesis testing is performed separately for each different value of the $a$-boson mass. Nuisance parameters describing systematic uncertainties are included by following the prescription in Ref.~\cite{Cranmer:2014lly} and have Gaussian constraint terms. Finally, a dedicated set of nuisance parameters is introduced to describe Poisson fluctuations in the background yields estimated from simulation.
 
\begin{figure*}[!htbp]
\begin{center}
\subfloat[]{\includegraphics[width=0.48\textwidth]{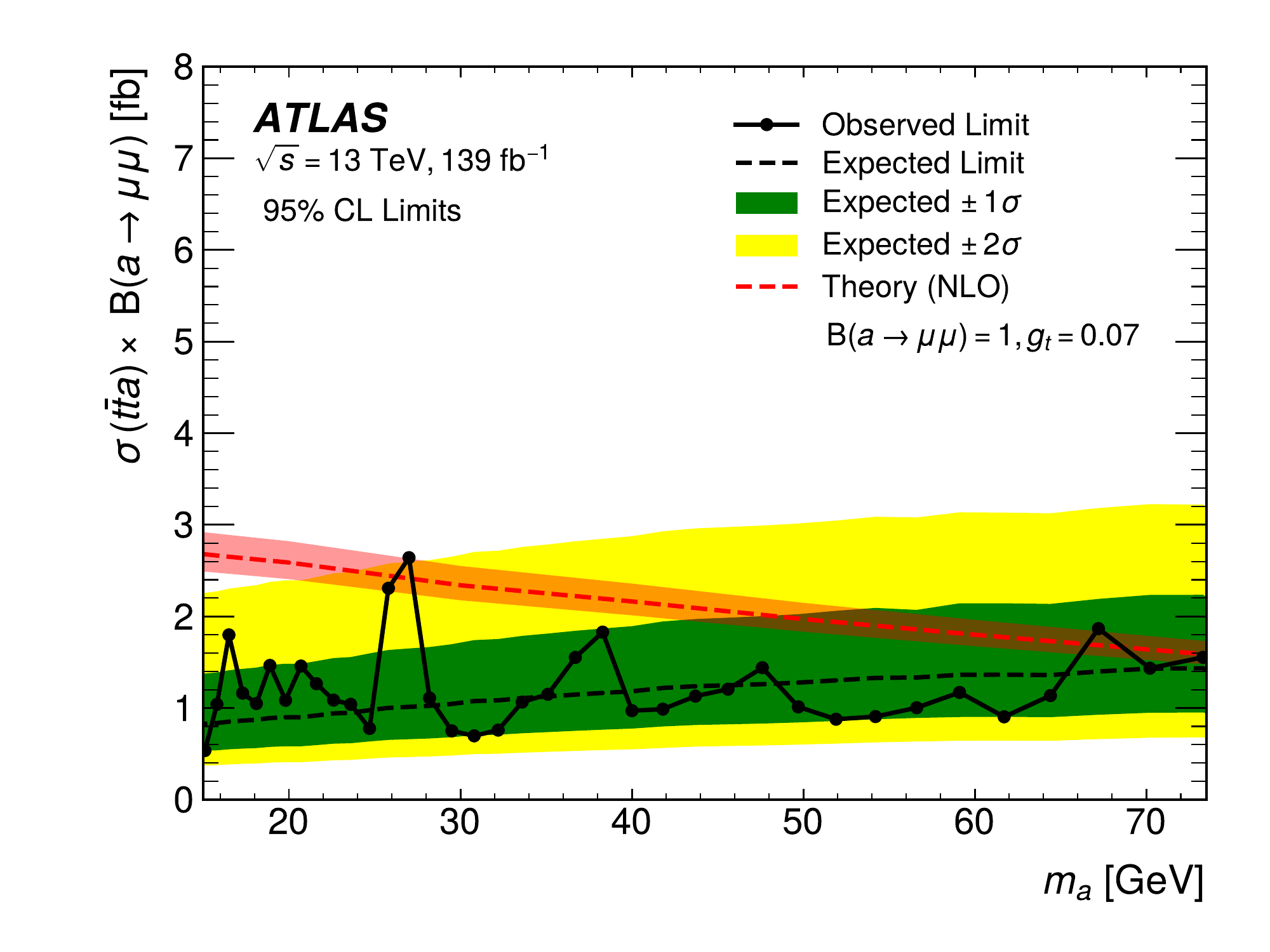}}
\subfloat[]{\includegraphics[width=0.48\textwidth]{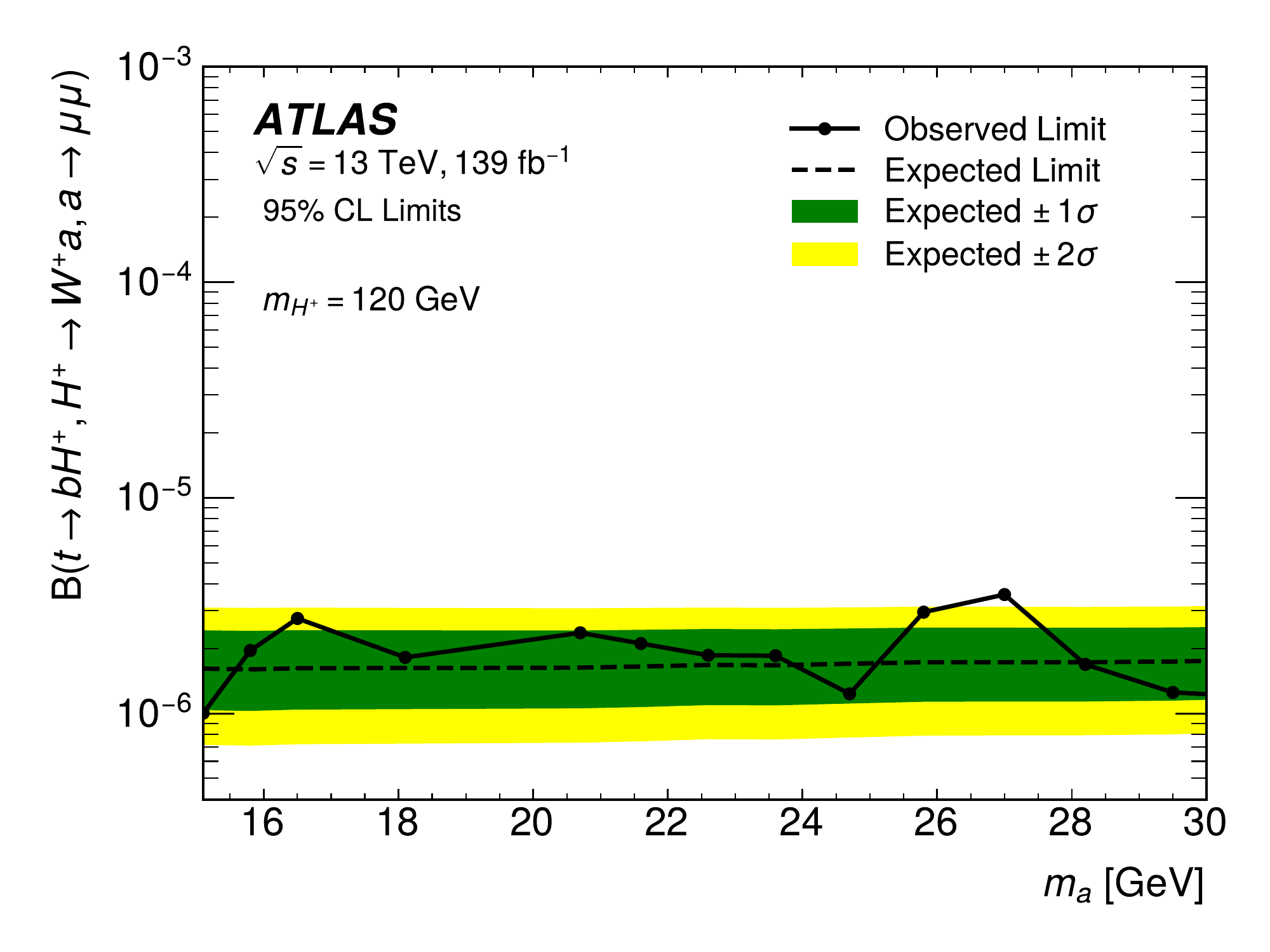}}\\
\subfloat[]{\includegraphics[width=0.48\textwidth]{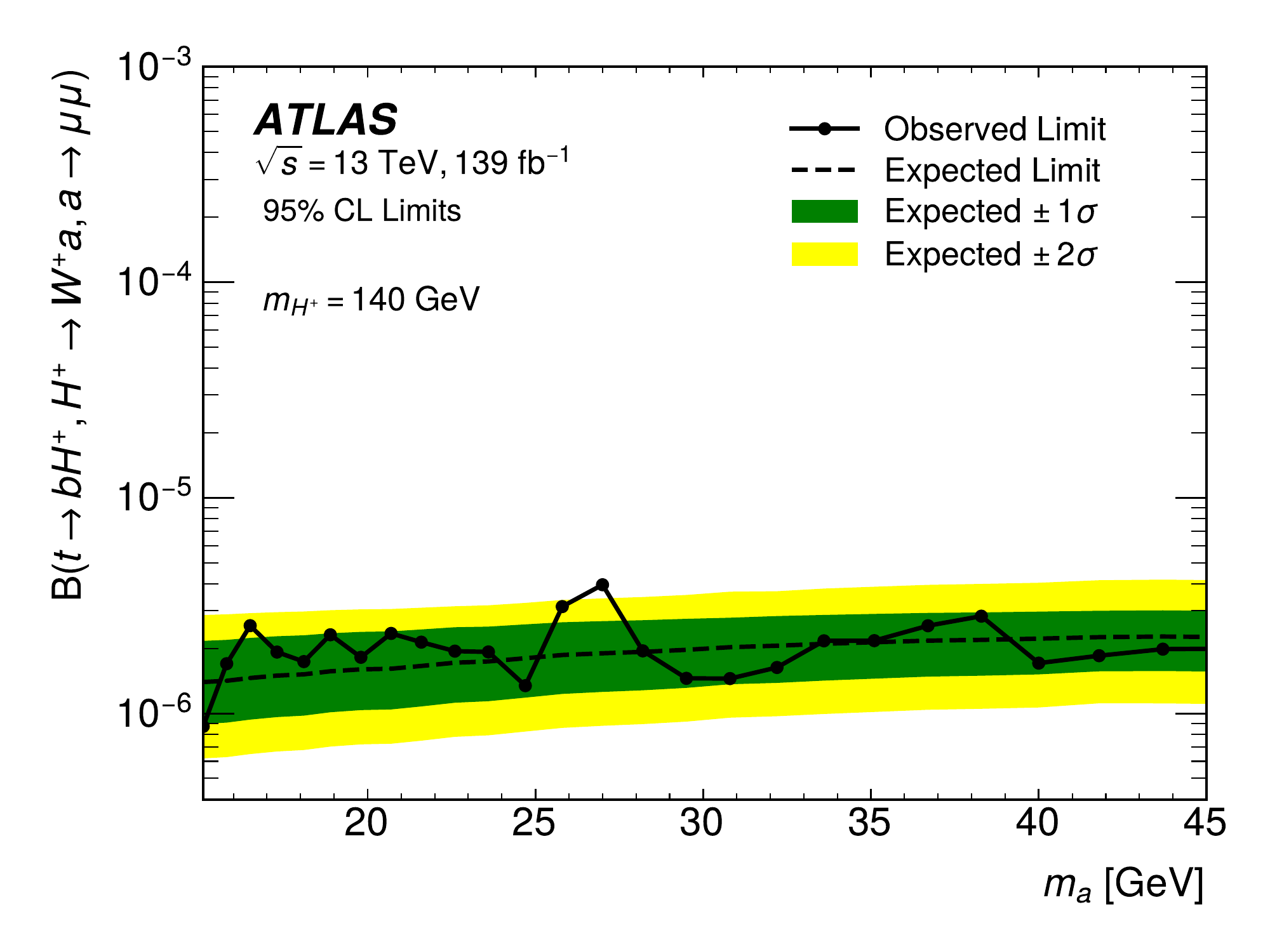}}
\subfloat[]{\includegraphics[width=0.48\textwidth]{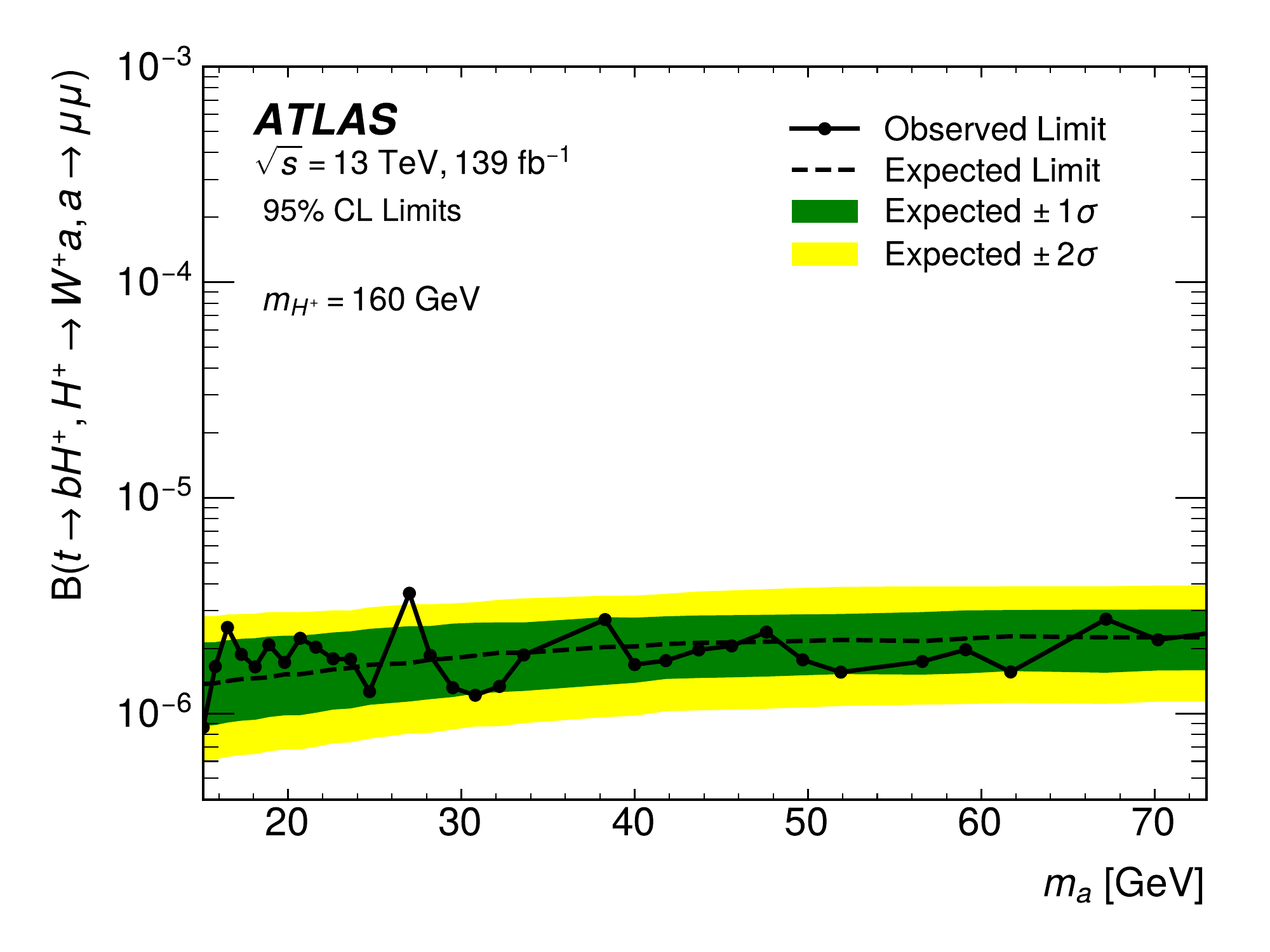}}\\
\caption{(a) Expected and observed 95\% confidence level (CL) upper limits on the signal cross section shown as a function of $m_a$ for $\tta$ production, and a comparison with the cross section predicted by the model described in the text. In (b), (c) and (d), expected and observed 95\% CL upper limits on the branching ratio $\text{B}(t\to bH^+, H^+ \to W^+a, a\to\mu\mu)$ are shown as a function of $m_a$ for $H^{+}\rightarrow W^{+}a$, assuming a top-pair cross section of $\sigma(pp\rightarrow \ttbar)= 833$~pb~\cite{Beneke:2011mq,Cacciari:2011hy,Baernreuther:2012ws,Czakon:2012zr,Czakon:2012pz,Czakon:2013goa,Czakon:2011xx}, for three values of the charged Higgs boson's mass: $m_{H^{+}} = 120$~\GeV, $m_{H^{+}} = 140$~\GeV, and $m_{H^{+}} = 160$~\GeV, respectively.}
\label{fig:Unblinded_cls}
\end{center}
\end{figure*}
 
As shown in Figure~\ref{fig:Unblinded_SR}, good agreement is observed between data and the expected background, suggesting the absence of a signal. A hypothetical $\tta$ signal with a cross section of $4\,\text{fb}$ at $m_a=35~\GeV$ is also depicted in Figure~\ref{fig:Unblinded_SR}, showing the typical concentrated excess that would have been expected from a signal.
No significant excess is observed, and the smallest local $p$-value is 0.008 at $m_a=27~\GeV$, corresponding to a local significance of about $2.4\,\sigma$. The slight excess at $m_a=27~\GeV$ is observed in both the \euu and \uuu channels. Upper limits on the signal cross section are determined using the CL$_\text{s}$ prescription~\cite{Read:2002hq,Cowan:2010js} and are shown in Figure~\ref{fig:Unblinded_cls}(a) together with a comparison with a prediction from a theoretical model for the $\tta$ analysis. The model assumes a single coupling $-iy_tg_t a(\bar{t}\gamma_5 t)/\sqrt{2}$, where $y_t$ is the top-quark Yukawa coupling, and the cross section is calculated at NLO precision in QCD. Only $a\to \mu\mu$ decays are included. A value of $g_t=0.07$ is chosen as an example in Figure~\ref{fig:Unblinded_cls}(a) since it approximately reproduces the largest cross section not excluded by this analysis. The uncertainty bands include both the scale and PDF variations. In 2HDM+$a$ models~\cite{Bauer:2017ota}, if the heavy Higgs bosons decouple, the coupling $g_t$ is given by $\sin\theta/\tan\beta$, where $\theta$ is the mixing angle between the light and heavy pseudoscalar bosons, and $\tan\beta$ is the ratio of the vacuum expectation values of the two scalar doublets.
 
\FloatBarrier
 
\section{Conclusion}
 
A search for a new pseudoscalar $a$-boson produced in events with a top-quark pair where the $a$-boson decays into a pair of muons, $\auu$, is performed. The analysis is based on the dataset of $pp$ collisions at $\sqrt{s}=13~\TeV$ recorded by the ATLAS detector in 2015 to 2018, corresponding to an integrated luminosity of $139~\fb^{-1}$. The search targets the final state where one top quark decays leptonically, resulting in two signal regions with three leptons, either $\euu$ or $\uuu$, and multiple jets. No significant excess of events above the Standard Model expectation is observed and interpretations in two signal models are considered. Upper limits are set on the cross section for associated production of $\ttbar$ and a new pseudoscalar, $pp \to \tta$, times the branching ratio $\text{B}(\auu)$ in the mass range $15~\GeV <m_a<72~\GeV$, and exclude signals with $\sigma(t\bar{t}a)\text{B}(\auu)$ above $0.5{-}3$~fb at 95$\%$ confidence level. The search also sets upper limits on the branching ratio $\text{B}(t\rightarrow bH^+,H^+\rightarrow W^+a,\auu)$ in the range $(0.9{-}3.9) \times 10^{-6}$ at 95$\%$ confidence level for $m_{H^+}$ between $120~\GeV$ and $160~\GeV$ and $m_a$ between $15~\GeV$ and $72~\GeV$. The results are largely dominated by statistical uncertainties and are expected to improve as more LHC collision data are collected.
 
\section*{Acknowledgements}

% The next lines are included from the .//acknowledgements/Acknowledgements.tex input file

We thank CERN for the very successful operation of the LHC, as well as the
support staff from our institutions without whom ATLAS could not be
operated efficiently.
 
We acknowledge the support of
ANPCyT, Argentina;
YerPhI, Armenia;
ARC, Australia;
BMWFW and FWF, Austria;
ANAS, Azerbaijan;
CNPq and FAPESP, Brazil;
NSERC, NRC and CFI, Canada;
CERN;
ANID, Chile;
CAS, MOST and NSFC, China;
Minciencias, Colombia;
MEYS CR, Czech Republic;
DNRF and DNSRC, Denmark;
IN2P3-CNRS and CEA-DRF/IRFU, France;
SRNSFG, Georgia;
BMBF, HGF and MPG, Germany;
GSRI, Greece;
RGC and Hong Kong SAR, China;
ISF and Benoziyo Center, Israel;
INFN, Italy;
MEXT and JSPS, Japan;
CNRST, Morocco;
NWO, Netherlands;
RCN, Norway;
MEiN, Poland;
FCT, Portugal;
MNE/IFA, Romania;
MESTD, Serbia;
MSSR, Slovakia;
ARRS and MIZ\v{S}, Slovenia;
DSI/NRF, South Africa;
MICINN, Spain;
SRC and Wallenberg Foundation, Sweden;
SERI, SNSF and Cantons of Bern and Geneva, Switzerland;
MOST, Taiwan;
TENMAK, T\"urkiye;
STFC, United Kingdom;
DOE and NSF, United States of America.
In addition, individual groups and members have received support from
BCKDF, CANARIE, Compute Canada and CRC, Canada;
PRIMUS 21/SCI/017 and UNCE SCI/013, Czech Republic;
COST, ERC, ERDF, Horizon 2020 and Marie Sk{\l}odowska-Curie Actions, European Union;
Investissements d'Avenir Labex, Investissements d'Avenir Idex and ANR, France;
DFG and AvH Foundation, Germany;
Herakleitos, Thales and Aristeia programmes co-financed by EU-ESF and the Greek NSRF, Greece;
BSF-NSF and MINERVA, Israel;
Norwegian Financial Mechanism 2014-2021, Norway;
NCN and NAWA, Poland;
La Caixa Banking Foundation, CERCA Programme Generalitat de Catalunya and PROMETEO and GenT Programmes Generalitat Valenciana, Spain;
G\"{o}ran Gustafssons Stiftelse, Sweden;
The Royal Society and Leverhulme Trust, United Kingdom.
 
The crucial computing support from all WLCG partners is acknowledged gratefully, in particular from CERN, the ATLAS Tier-1 facilities at TRIUMF (Canada), NDGF (Denmark, Norway, Sweden), CC-IN2P3 (France), KIT/GridKA (Germany), INFN-CNAF (Italy), NL-T1 (Netherlands), PIC (Spain), ASGC (Taiwan), RAL (UK) and BNL (USA), the Tier-2 facilities worldwide and large non-WLCG resource providers. Major contributors of computing resources are listed in Ref.~\cite{ATL-SOFT-PUB-2021-003}.

% End of text imported from the .//acknowledgements/Acknowledgements.tex input file

\clearpage

\printbibliography
 
\clearpage
 
\input{atlas_authlist}

\end{document}

%% file: atlas_authlist.tex
% ATLAS Collaboration author list
% Reference date of HDBS-2020-12 is 2023-03-27
% Author list last updated on date 19-DEC-23
% Data extracted on 19-Dec-2023 for paper reference HDBS-2020-12
% at 8:12am
 
\begin{flushleft}
\hypersetup{urlcolor=black}
{\Large The ATLAS Collaboration}

\bigskip

\AtlasOrcid[0000-0002-6665-4934]{G.~Aad}$^\textrm{\scriptsize 102}$,
\AtlasOrcid[0000-0002-5888-2734]{B.~Abbott}$^\textrm{\scriptsize 120}$,
\AtlasOrcid[0000-0002-1002-1652]{K.~Abeling}$^\textrm{\scriptsize 55}$,
\AtlasOrcid[0000-0001-5763-2760]{N.J.~Abicht}$^\textrm{\scriptsize 49}$,
\AtlasOrcid[0000-0002-8496-9294]{S.H.~Abidi}$^\textrm{\scriptsize 29}$,
\AtlasOrcid[0000-0002-9987-2292]{A.~Aboulhorma}$^\textrm{\scriptsize 35e}$,
\AtlasOrcid[0000-0001-5329-6640]{H.~Abramowicz}$^\textrm{\scriptsize 151}$,
\AtlasOrcid[0000-0002-1599-2896]{H.~Abreu}$^\textrm{\scriptsize 150}$,
\AtlasOrcid[0000-0003-0403-3697]{Y.~Abulaiti}$^\textrm{\scriptsize 117}$,
\AtlasOrcid[0000-0003-0762-7204]{A.C.~Abusleme~Hoffman}$^\textrm{\scriptsize 137a}$
% ...
\bigskip
\\

% ...
$^{29}$Physics Department, Brookhaven National Laboratory, Upton NY; United States of America.\\
% ...
$^{35}$$^{(a)}$Facult\'e des Sciences Ain Chock, R\'eseau Universitaire de Physique des Hautes Energies - Universit\'e Hassan II, Casablanca;$^{(b)}$Facult\'{e} des Sciences, Universit\'{e} Ibn-Tofail, K\'{e}nitra;$^{(c)}$Facult\'e des Sciences Semlalia, Universit\'e Cadi Ayyad, LPHEA-Marrakech;$^{(d)}$LPMR, Facult\'e des Sciences, Universit\'e Mohamed Premier, Oujda;$^{(e)}$Facult\'e des sciences, Universit\'e Mohammed V, Rabat;$^{(f)}$Institute of Applied Physics, Mohammed VI Polytechnic University, Ben Guerir; Morocco.\\
% ...
$^{49}$Fakult\"{a}t Physik , Technische Universit{\"a}t Dortmund, Dortmund; Germany.\\
% ...
$^{55}$II. Physikalisches Institut, Georg-August-Universit\"{a}t G\"ottingen, G\"ottingen; Germany.\\
% ...
$^{102}$CPPM, Aix-Marseille Universit\'e, CNRS/IN2P3, Marseille; France.\\
% ...
$^{117}$Department of Physics, New York University, New York NY; United States of America.\\
% ...
$^{120}$Homer L. Dodge Department of Physics and Astronomy, University of Oklahoma, Norman OK; United States of America.\\
% ...
$^{137}$$^{(a)}$Departamento de F\'isica, Pontificia Universidad Cat\'olica de Chile, Santiago;$^{(b)}$Millennium Institute for Subatomic physics at high energy frontier (SAPHIR), Santiago;$^{(c)}$Instituto de Investigaci\'on Multidisciplinario en Ciencia y Tecnolog\'ia, y Departamento de F\'isica, Universidad de La Serena;$^{(d)}$Universidad Andres Bello, Department of Physics, Santiago;$^{(e)}$Instituto de Alta Investigaci\'on, Universidad de Tarapac\'a, Arica;$^{(f)}$Departamento de F\'isica, Universidad T\'ecnica Federico Santa Mar\'ia, Valpara\'iso; Chile.\\
% ...
$^{150}$Department of Physics, Technion, Israel Institute of Technology, Haifa; Israel.\\
$^{151}$Raymond and Beverly Sackler School of Physics and Astronomy, Tel Aviv University, Tel Aviv; Israel.\\
% ...
$^{a}$ Also Affiliated with an institute covered by a cooperation agreement with CERN.\\
$^{b}$ Also at An-Najah National University, Nablus; Palestine.\\
$^{c}$ Also at Borough of Manhattan Community College, City University of New York, New York NY; United States of America.\\
$^{d}$ Also at Center for High Energy Physics, Peking University; China.\\
$^{e}$ Also at Center for Interdisciplinary Research and Innovation (CIRI-AUTH), Thessaloniki; Greece.\\
$^{f}$ Also at Centro Studi e Ricerche Enrico Fermi; Italy.\\
% ...

\end{flushleft}

% Created with Glance <Atlas.Glance@cern.ch>